\documentclass[11pt,letterpaper]{article}

\usepackage{graphicx}
\usepackage{float}
\usepackage{subcaption}
\usepackage{psfrag}
\usepackage{amsmath,amssymb}
\usepackage{url}
\usepackage[margin=1in]{geometry}
\usepackage[numbers]{natbib}
\usepackage{xcolor}
\usepackage{multirow}
\usepackage{tabularx}

\begin{document}
\renewcommand{\abstractname}{\vspace{-3\baselineskip}}
\title{A Utilization Model for Optimization of Checkpoint Intervals in Distributed Stream Processing Systems}
\author{Sachini~Jayasekara,
        Aaron~Harwood,
        and~Shanika~Karunasekera}
\date{%
   School of Computing and Information Systems, \\
   The University of Melbourne.\\%
    }

\maketitle
\begin{abstract}
\normalsize
State-of-the-art distributed stream processing systems such as Apache Flink and Storm have recently included checkpointing to provide fault-tolerance for stateful applications. This is a necessary eventuality as these systems head into the Exascale regime, and is evidently more efficient than replication as state size grows. However current systems use a nominal value for the checkpoint interval, indicative of assuming roughly 1 failure every 19 days, that does not take into account the salient aspects of the checkpoint process, nor the system scale, which can readily lead to inefficient system operation. To address this shortcoming, we provide a rigorous derivation of \emph{utilization} -- the fraction of total time available for the system to do useful work -- that incorporates checkpoint interval, failure rate, checkpoint cost, failure detection and restart cost, depth of the system topology and message delay. Our model yields an elegant expression for utilization and provides an optimal checkpoint interval given these parameters, interestingly showing it to be dependent only on checkpoint cost and failure rate. We confirm the accuracy and efficacy of our model through experiments with Apache Flink, where we obtain improvements in system utilization for every case, especially as the system size increases. Our model provides a solid theoretical basis for the analysis and optimization of more elaborate checkpointing approaches.
\end{abstract}

\section{Introduction}\label{sec:introduction}

The development of Exascale stream processing systems~\cite{6877346,MSU-CSE-06-2,1467855}, to cope with an ever increasing big data demand, presents many challenges, one of which is \emph{fault-tolerance} -- as the degree of parallelism in a stream processing system increases, the mean time to failure (MTTF) of the stream processing system as a whole decreases~\cite{5645453,1303239}. For instance, MTTF of Exascale systems is anticipated to be in minutes~\cite{phdthesis,doi:10.1177/1094342009347767} which arises the need for efficient fault tolerance approaches. Failures expected in stream processing systems can be of different types. A failure may occur in hardware, e.g. power supplies and memory read errors, that can lead to correlated failure of all processes on a machine, or may occur in the application, e.g. as an exception, where a single process or thread on the machine fails. In this work we consider the system as a whole with a single failure rate prescribed to it. Failures may occur at any time, even while the system is trying to recover from a previous failure. Without loss in generality our results are thereby oblivious to the exact location of the failure.

The latest practical deployments of distributed stream processing systems, such as \emph{Apache Flink} and \emph{Storm}~\cite{Carbone2015ApacheFS,Toshniwal:2014:STO:2588555.2595641} which typically exhibit a directed acyclic graph (DAG) architecture of stream processing operators, have recently included \emph{checkpointing}~\cite{Carbone:2017:SMA:3137765.3137777,Low:2012:DGF:2212351.2212354} to increase fault-tolerance for the case when stream operators are stateful~\cite{CastroFernandez:2013:ISO:2463676.2465282,5958214}. The general approach employed, called global or \emph{system-wide checkpointing}, is to only buffer messages at the sources of the topology and to replay such messages when the entire topology rolls back to a checkpoint. This avoids the problem of buffering messages at every operator of the topology (local checkpointing) which would otherwise become infeasible with high volume data streams; though global checkpointing requires the entire topology to roll-back on failure which also has limitations. While stream processing systems were originally envisaged to use only stateless operators, the use of stateful operators has grown to accommodate a greater range of complex stream processing such as large graph processing~\cite{7498275,Carbone2015ApacheFS}, machine learning~\cite{7891628,Meng:2016:MML:2946645.2946679} and general parallel processing~\cite{8622312}; where undertaking these computations ``in stream" is known to be a more effective approach than offloading to a third party system~\cite{8622147}. Checkpointing is argued as a more efficient alternative to state replication~\cite{7927721,8025331,7164919,6267921}, which becomes more evident as the state size grows.

In this paper we address the fundamental checkpointing question: how often should checkpoints be computed? We answer this question by formulating an analytical expression for \emph{utilization} -- the fraction of total time available to do useful work -- by which we can express the optimal checkpoint interval as that which maximizes utilization, in the context of a stream processing system as a whole, accurately modeling the state-of-the-art implementations. We provide experimental results with Apache Flink that show the efficacy of our analytical model for real world improvements in utilization, compared to the common use of default checkpoint interval settings. Our derivation is similar to but more appropriate for stream processing than the seminal work of Daly~\cite{Daly:2006:HOE:1134241.1708449,Daly:2003:MPO:1757599.1757601}, which follows from Young~\cite{Young:1974:FOA:361147.361115} who introduced a first order model to approximate the optimal checkpoint interval, where we also take into account higher order effects such as failure during recovery and multiple failures during a single interval. Furthermore Daly's approach minimizes the total runtime of the system, given a bounded workload, whereas for stream processing systems the workload is unbounded and so our utilization model is more applicable. The very recent work of Zhuang et al.~\cite{8487327} proposes a checkpointing model for stream processing systems, also based on Daly, that is workload aware. They do not consider failure during recovery, they make some assumptions on average time to a failure within a checkpoint period (similar to Daly's first order model) and they do not show any experiments to confirm their results.

This paper gives some additional related work in Section~\ref{sec:relatedWork}, provides a fundamental derivation of utilization in Section~\ref{sec:utilizationModel}, applies this to a stream processing system in Section~\ref{sec:utilizationStream}, shows experimental results in Section~\ref{sec:modelEvaluation} and some concluding remarks in Section~\ref{sec:conclusion}.

\section{Related Work}
\label{sec:relatedWork}

Checkpointing is a widely used fault tolerance and state management mechanism which is being used in existing data/stream processing systems. Several approaches have been suggested to determine the optimal checkpointing frequency for an application using application specific parameters. Young~\cite{Young:1974:FOA:361147.361115} introduced a model to find the checkpointing frequency that minimizes the time wasted due to failures. However, this model does not take into account failures during recovery and checkpointing. Daly~\cite{Daly:2006:HOE:1134241.1708449,Daly:2003:MPO:1757599.1757601} improved this model and used a cost function to determine the frequency that gives the minimum total wall clock time to complete an application. The wall clock time consists of time spent on actual computations, checkpointing time, rework time and restart time. The model proposed by Jin et al.~\cite{5599253} for HPC environments requires sequential workload of the application. However, for streaming systems determining the time to complete an application is somewhat irrelevant since the workload is unbounded.

Naksinehaboon et al.~\cite{article} propose approaches focusing on reducing the wasted time of a system and Ling et al.~\cite{936236} and Ozaki et al.~\cite{1632007} propose approaches based on calculus of variations to approximate the optimal checkpointing frequency. Rahman et al.~\cite{7515706} also provide a model for volunteer computing environments which minimizes the completion time. This approach assumes that a faulty process can only start after a checkpoint interval and time to detect failure and restart is negligible, which is not the case with streaming applications which we show in our evaluations.

More elaborate checkpointing approaches such as multi-level checkpointing~\cite{7442859,6494566} have been introduced to further reduce the checkpointing overhead. Multi-level checkpointing considers multiple failure types and each failure type has a different checkpoint cost and a restart cost. In this approach, less expensive checkpoint types are performed more frequently while more expensive and more resilient checkpoint types are performed less frequently. For example, Mohror et al.~\cite{6494566} present a Markov model for a multi-level checkpointing system while Sheng Di et al.~\cite{7442859} suggest a pattern based two-level checkpointing model for HPC environments which assumes that failures do not occur during the recovery period. However, more work needs to be done on how we can adapt multi-level checkpointing to a stream processing system and how different factors such as depth of the system topology and message delay will influence multi-level checkpointing process in a streaming application.
 
Fialho et al.~\cite{5961713} propose a model for uncoordinated checkpointing where each processor performs checkpoints independently, but this is not the case for most streaming systems. Zhuang et al.~\cite{8487327} propose an optimal checkpointing model for stream processing applications where each operator has an independent checkpointing interval. However, this cannot be applied to existing stream processing systems such as Apache Storm and Flink which require a unique checkpointing interval for the application, not for each operator. Furthermore, for large-scale applications running on large clusters, as the number of nodes in the cluster increases, the time between failures can reduce from hours to minutes. Therefore, assumptions made on existing models, such as the sum of the checkpoint interval and checkpoint cost being significantly less than the mean time between failures~\cite{8487327} may no longer be valid.

Checkpointing is widely used in real world stream processing systems to support fault tolerance. Apache Flink~\cite{Carbone:2017:SMA:3137765.3137777}, a well-known stream processing system performs periodic checkpoints using a token-based mechanism. In this approach, a token is sent from source operators to the succeeding operators in the application and at the arrival of the token each operator performs the checkpoint and sends the token to its output operators. For cases where an operator has multiple inputs, it waits for the token from all the inputs to start the checkpoint. Apache Storm follows a similar approach for checkpointing and the token is referred to as \emph{checkpoint tuple}. A checkpoint is considered as completed only when all operators in the application complete their individual checkpoints.

\section{Fundamental Utilization Model}
\label{sec:utilizationModel}

In this section we provide a derivation of our utilization model from first principles, which considers an underlying abstract system, progressively refined with the salient aspects of checkpointing and recovery. In tandem we compare our utilization model with the outcomes of a stochastic simulation of the abstract system, to support the model's correctness. In Section~\ref{sec:utilizationStream} we extend this to a stream processing system and in Section~\ref{sec:modelEvaluation} we use real-world measurements from experiments using Apache Flink, which has an equivalent checkpointing process to our abstract system, to show the efficacy of our model for system optimization.

At a highlevel, we define the \emph{utilization}, $0 \leq U\leq 1$, of a system as the fraction of the system's time for which its resources are available to do useful work (to process load), as opposed to work done solely to maintain the system's operation, sometimes called overhead, which in our definition includes the overhead associated with loss and recovery from system failure -- in fact in this paper we assume this as the only significant source of overhead. The work done by the system to create a checkpoint, and the work done by the system from the checkpoint time to the time taken to detect an occurred failure and successfully restart from the checkpoint is not useful work under our definition and thereby detracts from the utilization. The only useful work is therefore the work done, without failure, between two consecutive checkpoints (not including the work to create the checkpoints), or between a successful restart to the next checkpoint. For systems that process streaming data, where the load varies over time, the utilization is an expression of idle time plus time doing work when load is present. If there is no load on the system then the utilization is wholely the idle time of the system.

\subsection{Without failure}

\begin{figure}[h]
\centering
\includegraphics[width=3in]{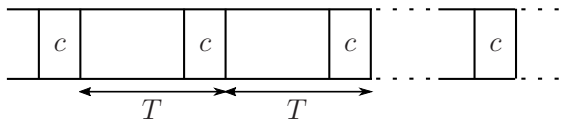}
\caption{Checkpointing with period $T$, checkpoint overhead $c$ and no failures occurring.}
\label{fig:noFaiiure}
\end{figure}

Consider a system that does checkpointing with a constant periodicity of $T$ seconds, i.e. if a single failure were to occur then less than $T$ seconds of work is lost.
Let $0\leq c\leq T$ be an abstract constant cost for checkpointing, here expressed without loss in generality as to how or when the checkpoint is created, in units of time, which without failure as shown in Fig.~\ref{fig:noFaiiure} leads to an expression for utilization:
\begin{equation}
\label{eq:usimple}
U=\frac{T-c}{T}.
\end{equation}

In our work we depict the checkpoint as being created in the last $c$ seconds of the period $T$, however mathematically it is immaterial as to exactly where in the period $T$ the system spends time creating the checkpoint, suffice to say that it completes exactly at the end of the period. The actual value of $c$ depends on the system itself (e.g. the performance of the disk drives for storing the checkpoint) and the application (how much data needs to be checkpointed) and we assume that this value is known or measured as needed (perhaps periodically measured if $T$ is to be periodically updated). Therefore the question to answer is what should be the value of $T$: too small (close to $c$) and the utilization is degraded due to spending too much time doing (unnecessary) checkpointing, too large and the utilization will be degraded due to too much work lost when failure occurs, as we show next.

\subsection{With failure and negligible restart cost} \label{sec:failures}

\begin{figure}[h]  
   \centering
	\includegraphics[height=1in]{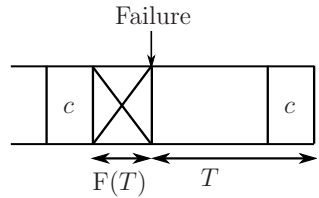}
	\caption{Failure resulting in lost time $\mathrm{F}(T)$ and instantaneous recovery followed by a successful period $T$.}
	\label{fig:1Failure}
\end{figure}

We model failure as a series of independent failure events having an exponential inter-arrival time given by random variable $\mathbf{X}$ with parameter $\lambda$, where the probability of failure at time $t$ is $\mathbb{P}[\mathbf{X}=t]=\lambda e^{-\lambda t}$, the probability of failure by time $t$ is $\mathbb{P}[\mathbf{X}<t]=1-e^{-\lambda t}$, and the mean time to failure is \[\mathbb{E}[\mathbf{X}]=\int_0^\infty t\,\mathbb{P}[\mathbf{X}=t]\,dt = \frac{1}{\lambda}.\] Assume that the system has just completed a checkpoint period. Either no failure happens within the next time period $T$, with probability $e^{-\lambda T}$, in which case the system successfully computes the next checkpoint, or failure happens within time $T$ as indicated in Fig.~\ref{fig:1Failure} and the system needs to restart from the existing checkpoint. In the later case we should ask at what point within time $T$ did the failure happen, more generally, given that failure does happen within time $T$ then what is the mean time to failure, $\mathrm{F}(T)$? To do this we normalize the distribution over the interval $[0,T]$ and re-evaluate the mean:
\begin{equation}\label{eq:meantime}
\begin{aligned}
\mathrm{F}(T)=\mathbb{E}[\mathbf{X}|\mathbf{X}<T]= \frac{\int_0^T t\,\mathbb{P}[\mathbf{X}=t]\,dt}{\mathbb{P}[\mathbf{X}<T]} \\
=\frac{{\mathrm{e}}^{T\,\lambda }-T\,\lambda-1}{\lambda \,({\mathrm{e}}^{T\,\lambda }-1)}<\frac{1}{\lambda}.
\end{aligned}
\end{equation}
The value for $\mathrm{F}(T)$ tells us the average amount of time lost (not utilized) if a failure happens, not having restarted yet. In this section we assume the time to detect the failure and to restart is negligible. Fig.~\ref{fig:1Failure} depicts only a single failure occurring preceding a successful period. However in general, the number of consecutive failures that could occur, $k$, before a successful period is unbounded and selected from the set $k\in\{0,1,2,\dotsc\}$ at random with a geometric distribution, $(1-p)^kp$, having parameter \[p=p_T = \mathbb{P}[\mathbf{X}\geq T]=1-\mathbb{P}[\mathbf{X}<T],\] giving an average number of consecutive failures \[\frac{1-p_T}{p_T}.\] Each consecutive failure looses on average an additional $\mathrm{F}(T)$ time (again not including restart and recovery time). We include this lost time into our utilization model from Eq.~\ref{eq:usimple} by expressing the \emph{effective} period, 
\[T_{eff}=T+\frac{1-p_T}{p_T}\mathrm{F}(T),\] and writing:
\begin{equation}
\label{eq:ufailure}
U=\frac{T-c}{T_{eff}}=\frac{\lambda \,\left(T-c\right)}{{\mathrm{e}}^{T\,\lambda }-1}.
\end{equation}

\subsection{Including the time to detect and recover from failure} \label{sec:recover}

\begin{figure}[h]   
        \centering 
		\includegraphics[height=1in]{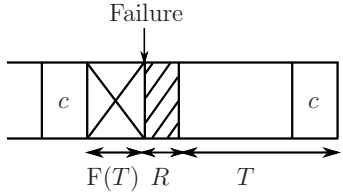}
		\caption{Failure resulting in lost time $\mathrm{F}(T)$ and recovery time $R$ followed by a successful period $T$.}
		\label{fig:R}
\end{figure}

This section includes the time to detect and recover from the failure, $R$, as shown in Fig.~\ref{fig:R}. In this case it is at first glance tempting to write \[T_\mathit{eff}=T+\frac{1-p_T}{p_T}(\mathrm{F}(T)+R),\] since every failure requires a restart, however we note that failure may also occur \emph{during} the restart, in which case we assume that the restart must itself start again. Similarly to the number of consecutive failures, the number of attempts to restart, $r\in\{1,2,\dotsc\}$, is selected at random using a geometric distribution (note that following a failure at least 1 restart is always required), $(1-p)^{r-1}p$, with parameter $p=p_R=\mathbb{P}[\mathbf{X}\geq R]$, leading to an average number of restarts $\frac{1}{p_R}\geq 1$. For any given failed restart attempt, given that we know the failure occurred within the restart time $R$ we know from Eq.~\ref{eq:meantime} that the average time lost is $\mathrm{F}(R)$. Taking all of this into account leads to
\[
T_\mathit{eff}=T+\frac{1-p_T}{p_T}\bigg(\mathrm{F}(T)+R+\Big(\tfrac{1}{p_R}-1\Big)\mathrm{F}(R)\bigg)
\]
and finally:
\begin{equation}
U=\frac{T-c}{T_\mathit{eff}}=\frac{\lambda \,\left(T-c\right)}{{\mathrm{e}}^{\lambda \,\left(R+T\right)}-{\mathrm{e}}^{R\,\lambda }}.
\label{eq:simpleSingleU}
\end{equation}
This completes the salient features of our checkpoint and restart system model for a single process. We adapt this to a distributed stream processing system in Section~\ref{sec:utilizationStream}.

\subsection{Optimization of utilization}

\begin{figure}[h]  
  \centering
  \includegraphics[width=2.6in]{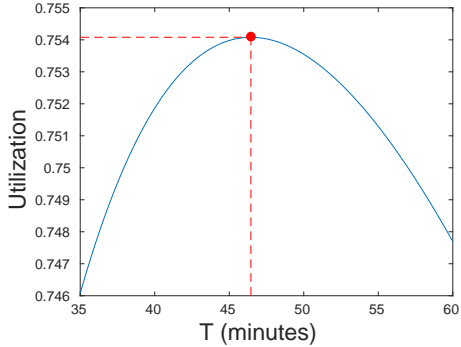}
   \caption{Utilization for $\lambda=0.005$ per minute, $c=5$ minutes, $R=10$ minutes.} 
   \label{fig:singleUEx}
\end{figure}

The value of $T$ that maximizes the utilization $U$ from Eq.~\ref{eq:simpleSingleU}, $T^*$, is found by solving $\frac{\partial U}{\partial T}=0$ for $T$:
\[
 \frac{\partial U}{\partial T} = \frac{\lambda }{{\mathrm{e}}^{\lambda \,\left(R+T\right)}-{\mathrm{e}}^{R\,\lambda }}-\frac{\lambda ^2\,{\mathrm{e}}^{\lambda \,\left(R+T\right)}\,\left(T-c\right)}{{\left({\mathrm{e}}^{\lambda \,\left(R+T\right)}-{\mathrm{e}}^{R\,\lambda }\right)}^2}=0,\]
\[T^* = \frac{c\,\lambda +{\mathrm{W}}\left(-{\mathrm{e}}^{-c\,\lambda -1}\right)+1}{\lambda },
\]
where $\mathrm{W}(z)$ is the Lambert $W$ function on the principal branch. Interestingly $T^*$ is not a function of $R$. For example, Fig.~\ref{fig:singleUEx} shows the utilization for varying $T$, with harsh values of $c$ and $R$, using the model expressed in Eq.~\ref{eq:simpleSingleU}. In this example, $U=0.7541$ is maximum when $T=T^*=46.452$ minutes.

\subsection{Comparison to stochastic simulation}
\begin{figure}[t]
    \centering
    \begin{subfigure}[h]{0.49\textwidth}
        \centering
        \includegraphics[width=2.5in, height=1.95in]{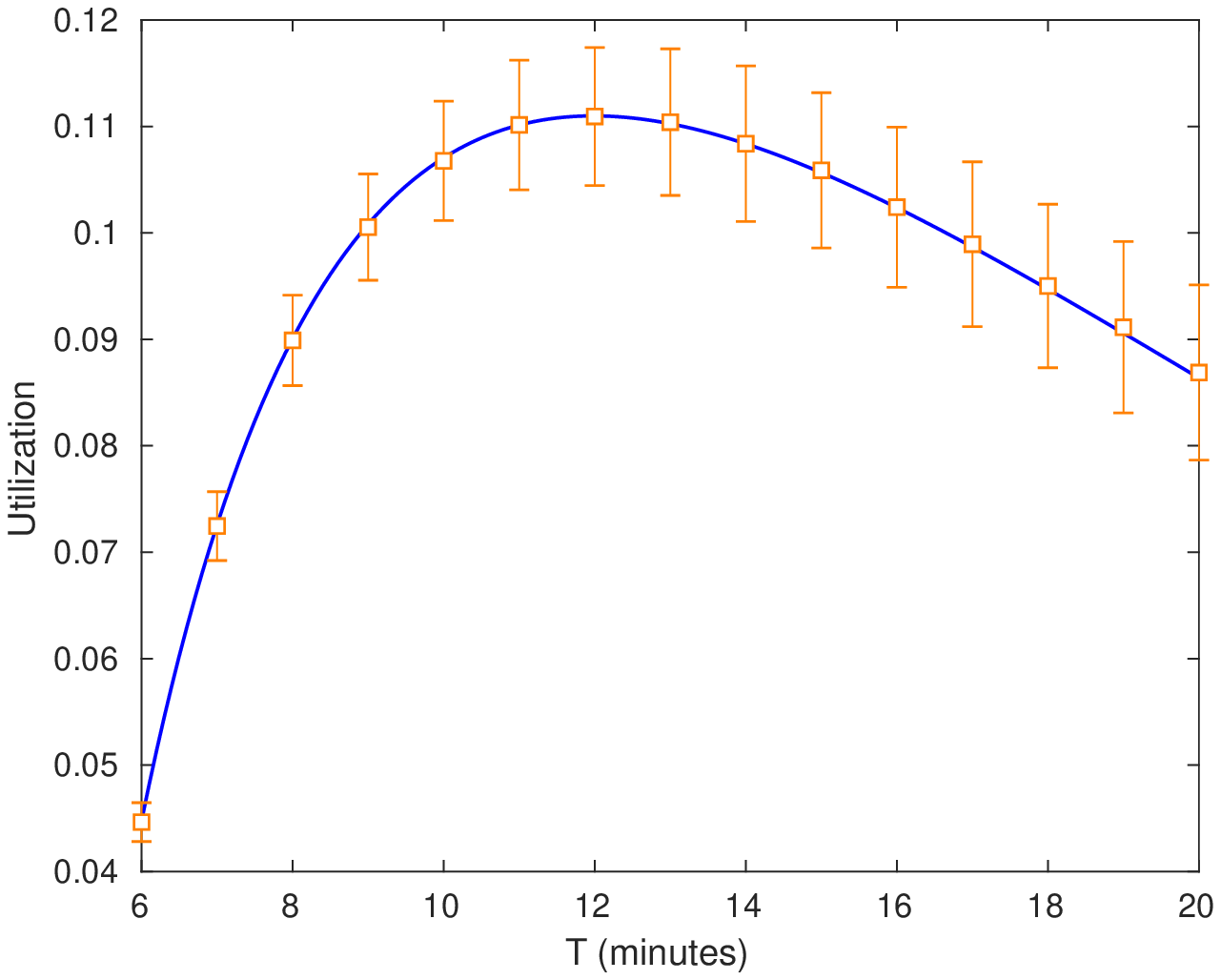}
        \caption{$\lambda$ = 0.1}    
        \label{fig:utiF_.1}
    \end{subfigure}
    \hfill
    \begin{subfigure}[h]{0.49\textwidth}  
        \centering 
        \includegraphics[width=2.5in, height=1.95in]{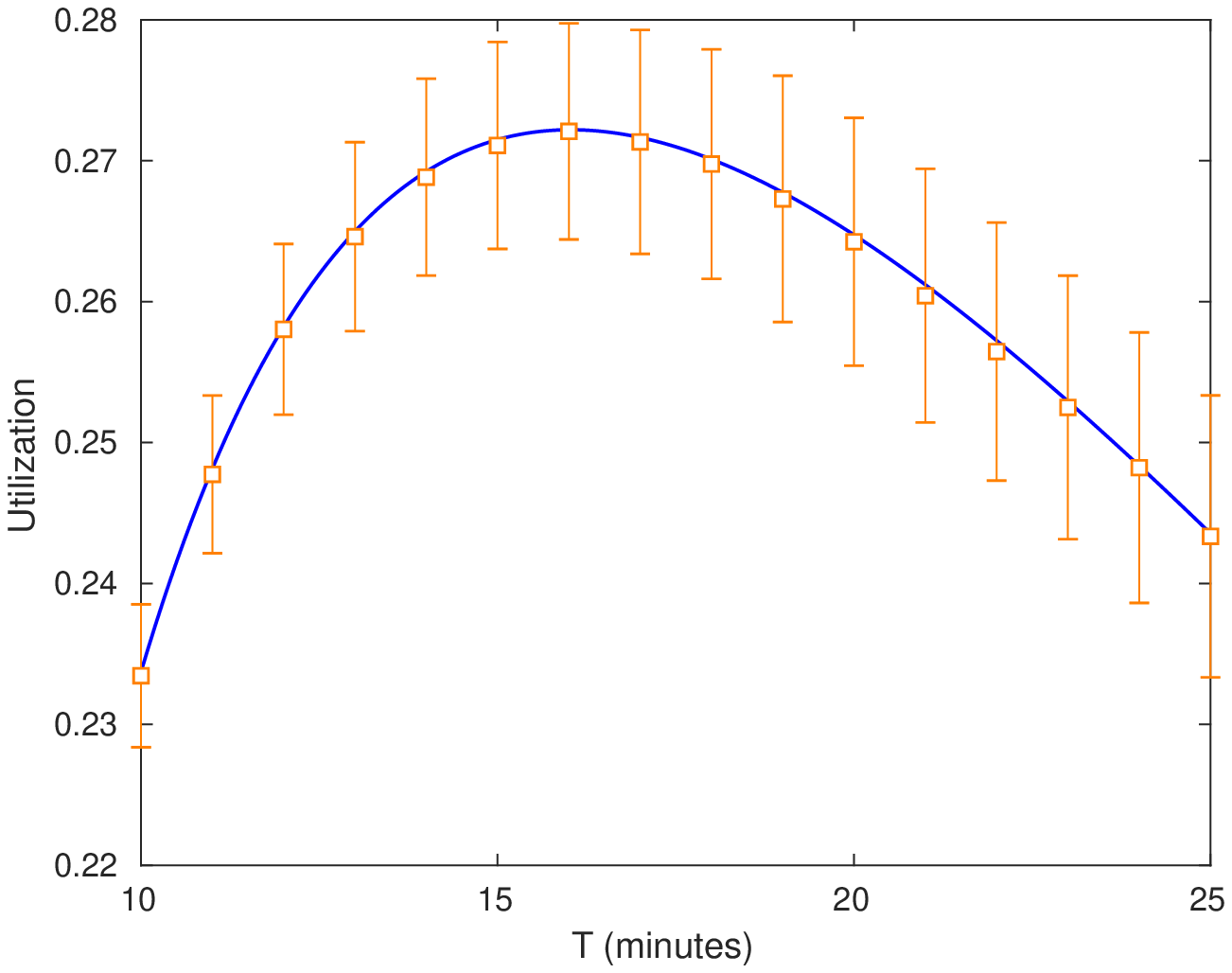}
        \caption{$\lambda$ = 0.05}     
        \label{fig:utiF_.05}
    \end{subfigure}
    \begin{subfigure}[h]{0.49\textwidth}   
        \centering 
        \includegraphics[width=2.5in, height=1.95in]{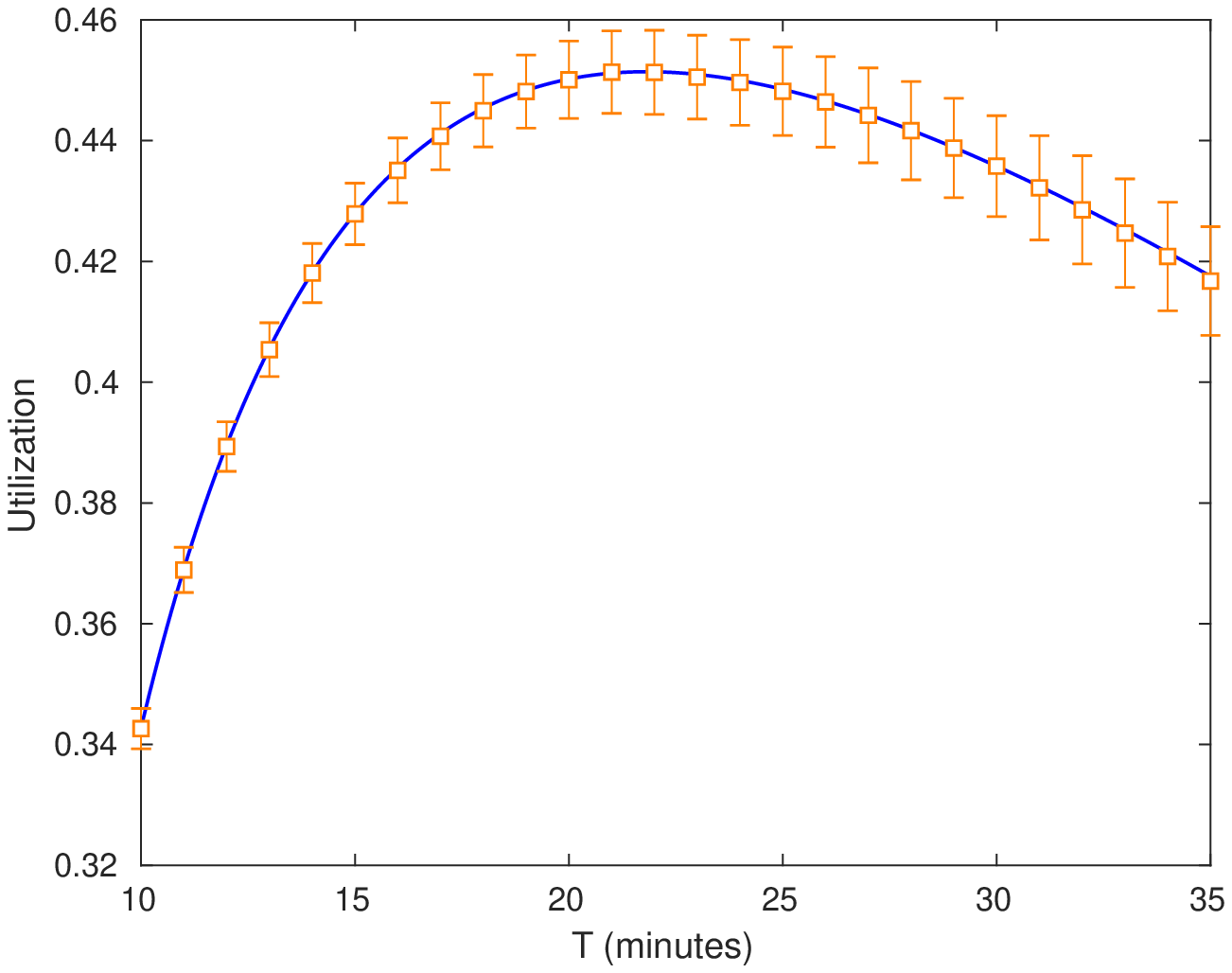}
        \caption{$\lambda$ = 0.025}     
        \label{fig:utiF_.025}
    \end{subfigure}
    \hfill
    \begin{subfigure}[h]{0.49\textwidth}   
        \centering 
        \includegraphics[width=2.5in, height=1.95in]{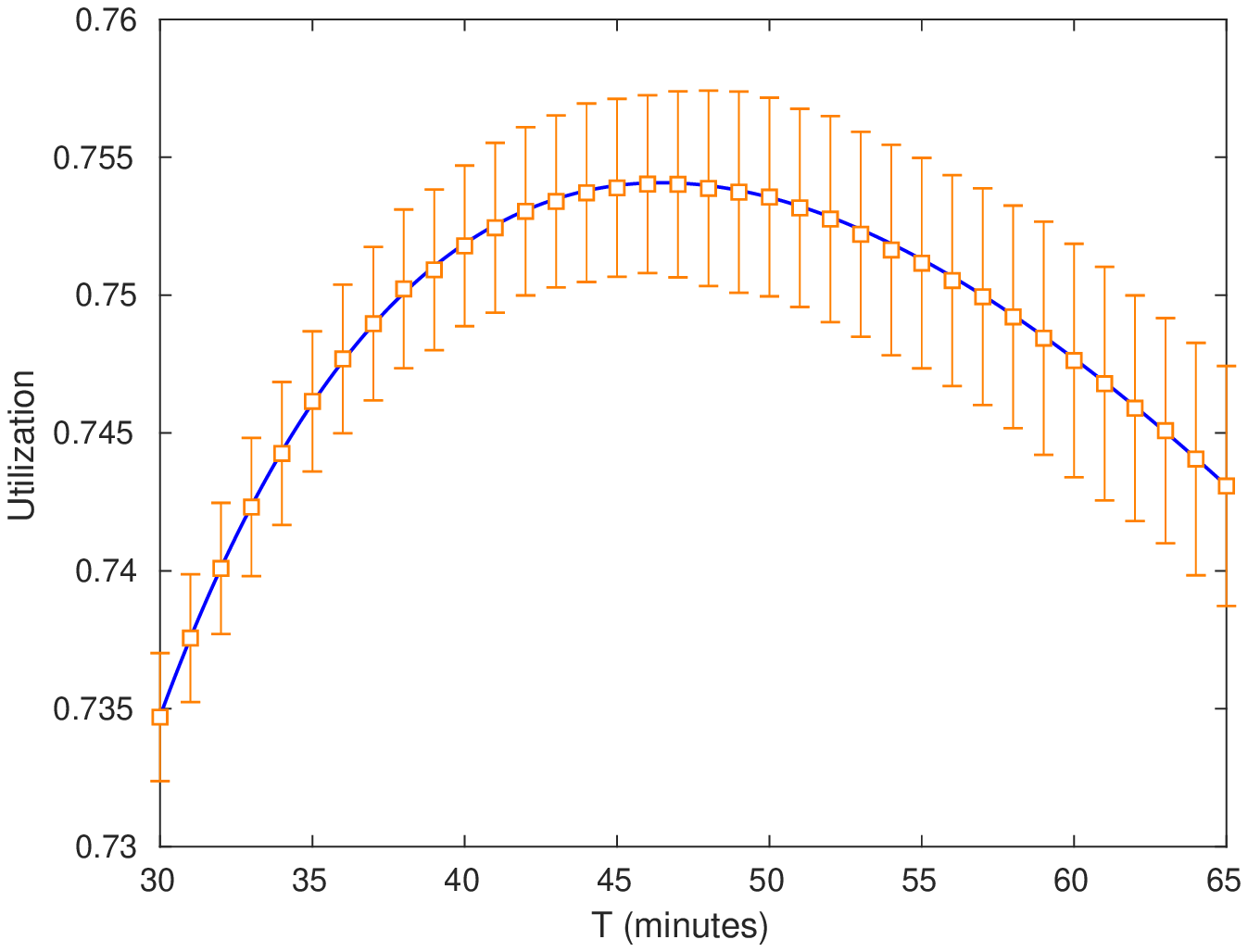}
        \caption{$\lambda$ = 0.005}      
        \label{fig:utiF_.005}
    \end{subfigure}
    \caption{Utilization for $R=10$ minutes $c=5$ minutes. Solid lines are theoretical utilization and data points with error bars are the average utilization and the standard deviation of simulation observations.} 
    \label{fig:expS}
\end{figure}

We developed a stochastic simulation to support the derived model. The simulation generates random failures based on an exponential distribution and calculates the utilization taking the randomly generated failures into account. Fig.~\ref{fig:expS} shows the utilization comparison between our model based on Eq.~\ref{eq:simpleSingleU} and the simulation results for different failure rates. The solid lines represent theoretical utilization and the data points with error bars represent the average utilization and the standard deviation of 250 runs of the simulation with each simulation running for $\frac{2000}{\lambda}$  minutes.

\section{Utilization of a Stream Processing System}
\label{sec:utilizationStream}

In this section we extend the proposed model to describe a distributed stream processing system, which is typically represented by a DAG of operators. An operator performs computations on its input stream(s) and may produce one or more output streams to succeeding operators in the DAG. Operators without input streams are sources and those without output streams are sinks. Token based checkpointing is the state-of-the-art approach used in stream processing systems. In this approach, a ``checkpoint token" is passed at regular checkpoint intervals from all sources of the DAG\footnote{If the system contains cycles then it can be logically transformed to a DAG for the purposes of checkpointing via the introduction of virtual sources at upstream points~\cite{Carbone:2017:SMA:3137765.3137777}.}, through the system covering all operators, to the sinks. The checkpoint token signals each operator to store its state as the next checkpoint. The \emph{system-wide} checkpoint completes when all operators have completed the checkpoint for a given token. At a failure event, all operators restore the last persisted checkpoint and continue processing. The system-wide checkpoint is the most common fault tolerance approach used in stream processing systems as it avoids the infeasible buffering of large volumes of data at every operator that would otherwise arise from high volume data streams. For example, well-known stream processing systems such as Storm, Flink and Samza use this method to support state management and fault tolerance. However, system-wide checkpointing requires the entire topology to rollback for each failure.

Some distributed checkpointing schemes such as Meteor Shower~\cite{6267921} have each operator checkpointing independently. 
This type of a scheme imposes additional overhead and needs more effort to maintain a consistent global state compared to system-wide checkpoint~\cite{articleGradvohl,10.1007/11827252_23}. For instance, this approach requires saving the message buffers at each operator to recover from failures whereas a system-wide checkpoint saves message buffers only at the sources. For a large-scale system dealing with large volumes of data buffering data at every operator is not desirable as it requires a significant amount of additional storage. 

\begin{figure}[h]  
   \centering
         \includegraphics[width=2.2in]{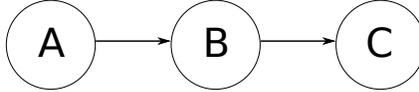}
   		\caption{DAG of operators representing a streaming application with 3 operators.} 
   		\label{fig:DAG}
\end{figure}

In streaming applications, operators can be of different degrees of parallelism residing in different nodes in a cluster. In such cases, for an application level failure affecting a single operator or for a hardware failure affecting a set of nodes in the cluster, the application undergoes the same recovery process despite the failure type. Fig.~\ref{fig:DAG} shows a DAG of three operators, $A$, $B$ and $C$. When determining the utilization, we assume that all operators are stateful and have identical checkpointing cost $0\leq c\leq T$ and identical cost to detect and recover from the failure, $R$.

\begin{figure}[h]  
        \centering
		\includegraphics[width=3.4in]{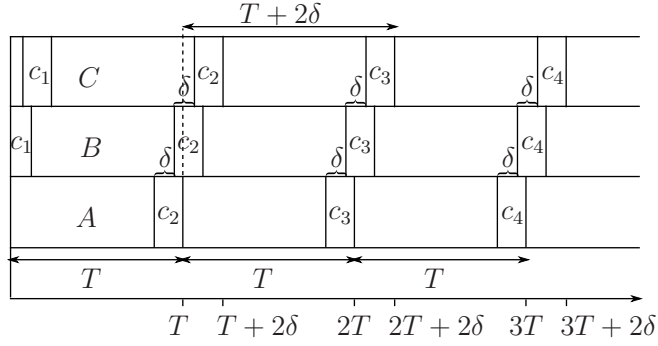}
		\caption{Checkpointing process of the streaming application represented in Fig.~\ref{fig:DAG}.} 
		\label{fig:streamTime}
    \end{figure}

\begin{figure}[h]  
        \centering 
		\includegraphics[width=3.3in]{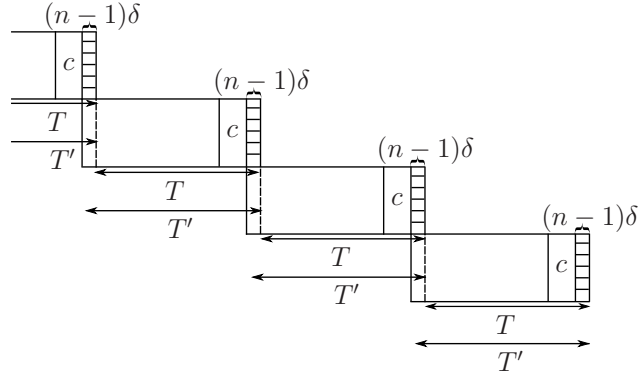}
		\caption{Alternative view of the checkpointing process of a streaming application showing the overlap between checkpoints and the computation.} 
		\label{fig:stream2}
\end{figure}

Fig.~\ref{fig:streamTime} depicts the operators from Fig.~\ref{fig:DAG} in terms of time and the flow of a checkpoint token that takes $\delta$ time to be transmitted from one operator to the next, causing the checkpoints for a given interval (note that we have numbered checkpoints $c_1, c_2, \dotsc$ for convenience) to be staggered. In this case the system-wide checkpoint is not complete until time $T+2\delta$. Let the longest path between all source and sink operators (critical path) in the DAG consist of $n$ operators including the source and the sink, then, without considering failure and using Fig.~\ref{fig:stream2}, we have \[T_\mathit{eff}=T^\prime= T+(n-1)\delta\] and so:
\begin{equation} 
U = \frac{T-c}{T_\mathit{eff}}=\frac{T-c}{T+(n-1)\delta}.
\label{eq:2}
\end{equation}

\subsection{Including failure and recovery cost}

In our model when a failure is detected in the stream processing system, all operators restore the last system-wide completed checkpoint, i.e. checkpoint $c_i$ for the largest $i$ where all operators have completed $c_i$. As discussed in Sections~\ref{sec:failures} and~\ref{sec:recover}, we include the time due to failures during $T^\prime$ and failures during $R$ to improve the accuracy of $T_\mathit{eff}$. Similarly to the number of consecutive failures, within time $T$ for a single operator, the number of attempts to complete $T^\prime$, $k^\prime\in\{1,2,\dotsc\}$, is selected at random using a geometric distribution, $(1-p)^{k^\prime}p$, with parameter \[p=p_{T^\prime}=\mathbb{P}[\mathbf{X}\geq T^\prime]=1-\mathbb{P}[\mathbf{X}<T^\prime],\] leading to an average number of consecutive failures \[\frac{1-p_{T^\prime}}{p_{T^\prime}}.\] Each consecutive failure looses on average an additional $\mathrm{F}(T^\prime)$ time. Similarly to the number of consecutive failures, the average number of restarts is $\frac{1}{p_R}\geq 1$ and the average time lost due to a failure during restart is $\mathrm{F}(R)$. Taking all of this into account leads to
\begin{equation}
\label{eq:streamteff1}
T_\mathit{eff}=T'+\frac{1-p_{T'}}{p_{T'}}\bigg(\mathrm{F}(T')+R+\Big(\tfrac{1}{p_R}-1\Big)\mathrm{F}(R)\bigg).
\end{equation}

\subsection{Including the overlap between consecutive checkpoints}

\begin{figure*}[h]
    \centering
    \begin{subfigure}[b]{0.48\textwidth}
        \centering
        \includegraphics[width=2in]{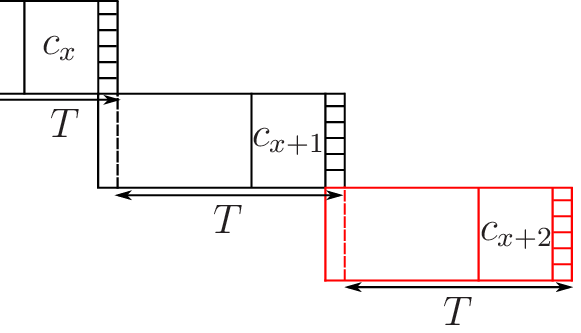}
        \caption{No failures}
        \label{fig:streamF1}
    \end{subfigure}
    \hfill
    \begin{subfigure}[b]{0.48\textwidth}
        \centering
        \includegraphics[width=2.2in]{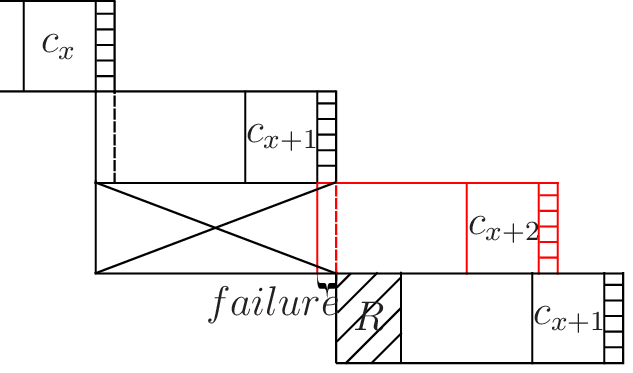}
        \caption{Failure between $0$ and $(n-1)\delta$ }
        \label{fig:streamF3}
    \end{subfigure}
    \vfill
    \begin{subfigure}[b]{0.9\textwidth}  
        \centering 
        \includegraphics[width=3.1in]{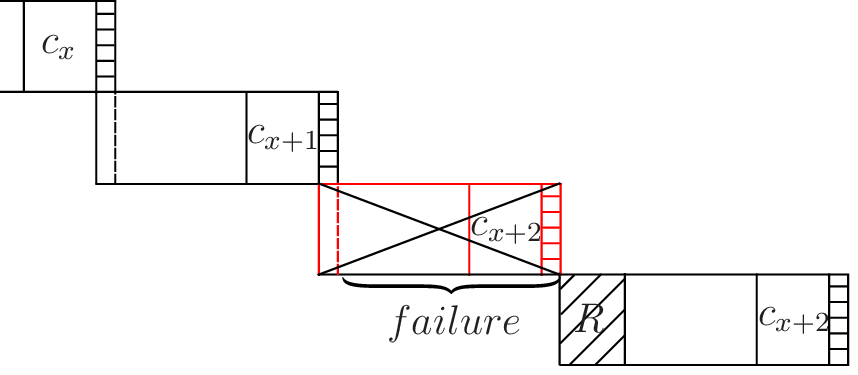}
        \caption{Failure between $(n-1)\delta$ and $T^\prime$}
        \label{fig:streamF2}
    \end{subfigure}
        \caption{Failure during $T^\prime.$} 
    \label{fig:streamF23}
\end{figure*}

During time $T^\prime$, the last completely persisted checkpoint changes as the computation progresses. For example, consider the computation interval between checkpoints $c_{x+1}$ and $c_{x+2}$ highlighted in Fig.~\ref{fig:streamF1}. If a failure occurs during $(n-1)\delta$ as shown in Fig.~\ref{fig:streamF3}, then the system has to restore from the checkpoint $c_{x}$ as checkpoint $c_{x+1}$ is still underway. If the failure occurs after $(n-1)\delta$, then the system can restore from the checkpoint $c_{x+1}$ as shown in Fig.~\ref{fig:streamF2}.

For $T^\prime$ leading up to checkpoint $c_x$, if a failure occurs during the first $(n-1)\delta$ time of the computation, then checkpoint restoration uses $c_{x-2}$. This is same as if the failure occurs during the previous $T^\prime$ leading up to checkpoint $c_{x-1}$. If a failure occurs between time $(n-1)\delta$ and $T^\prime$, then the checkpoint restoration is done using the previous checkpoint $c_x$. Therefore, recovery due to failure during the first $(n-1)\delta$ of a $T^\prime$ is the same as if the failure occurs during the previous $T^\prime$. The first $(n-1)\delta$ of $T^\prime$ is already represented by the previous $T^\prime$ as it overlaps with the previous $T^\prime$ as shown in Fig.~\ref{fig:stream2}. Hence, for one $T^\prime$, we only have to consider the time between $(n-1)\delta$ and $T^\prime$, ignoring the first $(n-1)\delta$ of $T^\prime$. 

The average number of consecutive failures before completing the first $(n-1)\delta$ of $T^\prime$ is \[\frac{1-p_{(n-1)\delta}}{p_{(n-1)\delta}},\] where \[p_{(n-1)\delta}=\mathbb{P}[\mathbf{X}\geq (n-1)\delta]=1-\mathbb{P}[\mathbf{X}<(n-1)\delta].\] Taking all of this into account, the effective period to complete $(n-1)\delta$ is:
\begin{multline}
\nonumber
(n-1)\delta + \frac{1-p_{(n-1)\delta}}{p_{(n-1)\delta}}\bigg(\mathrm{F}((n-1)\delta)+R+\Big(\tfrac{1}{p_R}-1\Big)\mathrm{F}(R)\bigg).
\end{multline}
Subtracting this from $T_\mathit{eff}$ in Eq.~\ref{eq:streamteff1} avoids the double representation of the first $(n-1)\delta$ and leads to:
\begin{multline}
\nonumber
T_\mathit{eff} =  T' + \frac{1-p_{T'}}{p_{T'}}\bigg(\mathrm{F}(T')+R+\Big(\tfrac{1}{p_R}-1\Big)\mathrm{F}(R)\bigg) - \\ 
\bigg((n-1)\delta + \frac{1-p_{(n-1)\delta}}{p_{(n-1)\delta}}\bigg(\mathrm{F}((n-1)\delta)+R+\Big(\tfrac{1}{p_R}-1\Big)\mathrm{F}(R)\bigg) \bigg)
\end{multline}
\begin{multline}
\nonumber
\qquad =  T + \frac{1-p_{T'}}{p_{T'}}\bigg(\mathrm{F}(T')+R+\Big(\tfrac{1}{p_R}-1\Big)\mathrm{F}(R)\bigg) - \\ 
\bigg(\frac{1-p_{(n-1)\delta}}{p_{(n-1)\delta}}\bigg(\mathrm{F}((n-1)\delta)+R+\Big(\tfrac{1}{p_R}-1\Big)\mathrm{F}(R)\bigg) \bigg).
\end{multline}
Finally:
\begin{equation}
U =  \frac{T-c}{T_\mathit{eff}}=\frac{\lambda \,{\mathrm{e}}^{\delta \,\lambda }\,\left(T-c\right)}{{\mathrm{e}}^{\lambda \,\left(R+T+\delta \,n\right)}-{\mathrm{e}}^{\lambda \,\left(R+\delta \,n\right)}}.\label{eq:DAGU}
\end{equation}
Remarkably the expression for utilization has a simple form that includes $\delta$ and $n$ in a natural way. This completes the salient features of our checkpoint and restart system model for a distributed stream processing system.

\subsection{Optimization of utilization}
The value of $T$ that maximizes the utilization $U$ from Eq.~\ref{eq:DAGU}, $T^*$, is found by solving $\frac{\partial U}{\partial T}=0$ for $T$:
\[
 \frac{\partial U}{\partial T} = -\frac{\lambda \,{\mathrm{e}}^{\delta \,\lambda 
 }}{{\mathrm{e}}^{\lambda \,\left(R+\delta \,\mathrm{n}\right)}-{\mathrm{e}}^{\lambda \,\left(R+T+\delta \,\mathrm{n}\right)}}\]
\[\qquad \qquad   -\frac{\lambda ^2\,{\mathrm{e}}^{\lambda \,\left(R+T+\delta \,\mathrm{n}\right)}\,{\mathrm{e}}^{\delta \,\lambda }\,\left(T-c\right)}{{\left({\mathrm{e}}^{\lambda \,\left(R+\delta \,\mathrm{n}\right)}-{\mathrm{e}}^{\lambda \,\left(R+T+\delta \,\mathrm{n}\right)}\right)}^2}=0
\]
\[ T^* = \frac{c\,\lambda +{\mathrm{W}}\left(-{\mathrm{e}}^{-c\,\lambda -1}\right)+1}{\lambda },
\]
where $\mathrm{W}(z)$ is the Lambert $W$ function on the principal branch. Interestingly $T^*$ is identical to that of a single process, i.e. it is independent of $n$ and $\delta$. 

\begin{figure}[H]  
    	\centering
        \includegraphics[width=2.2in]{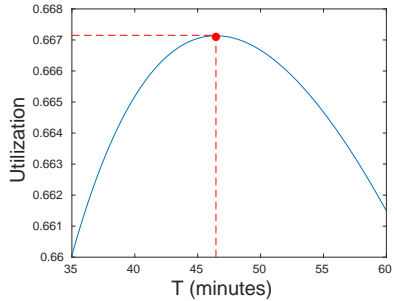}
   		\caption{Utilization of a DAG of operators for $\lambda=0.005$ per minute, $c=5$ minutes, $R=10$ minutes, $n=50$, $\delta=0.5$ minutes.} 
   		\label{fig:DAGUEx}
         \end{figure}

\begin{figure}[h]   
        \centering 
        \includegraphics[width=2.5in]{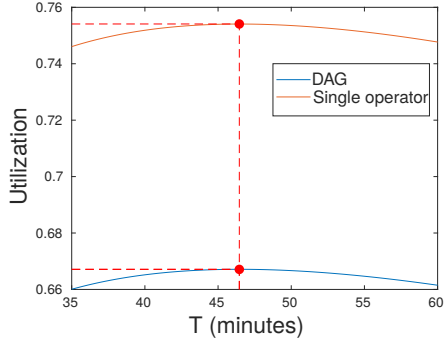}
        \caption{Utilization of a single operator and a DAG of operators for $\lambda=0.005$ per minute, $c=5$ minutes, $R=10$ minutes, $n=50$, $\delta=0.5$ minutes.} 
        \label{fig:dagvs1}   
\end{figure}

Fig.~\ref{fig:DAGUEx} shows the utilization for different values of $T$, using the model expressed in Eq.~\ref{eq:DAGU}. In this example, $U=0.667$ is the maximum when $T=46.452$ minutes. Fig.~\ref{fig:dagvs1} shows the utilization of a single operator and a DAG of operators for the same $\lambda,R,c$ values. As indicated in the figure, $T^*$ is identical for both cases despite the $\delta$ and $n$ terms introduced. However, the utilization is significantly less due to the impact of $\delta$ and $n$. For example, in the figure we can observe 11.6\% percentage decrease in utilization for a DAG of operators with $n=50$ compared to a single operator.

\subsection{Comparison to stochastic simulation}

\begin{figure}[t]       
    \centering
    \begin{subfigure}[b]{0.47\textwidth}
        \centering
        \includegraphics[width=2.5in, height=1.9in]{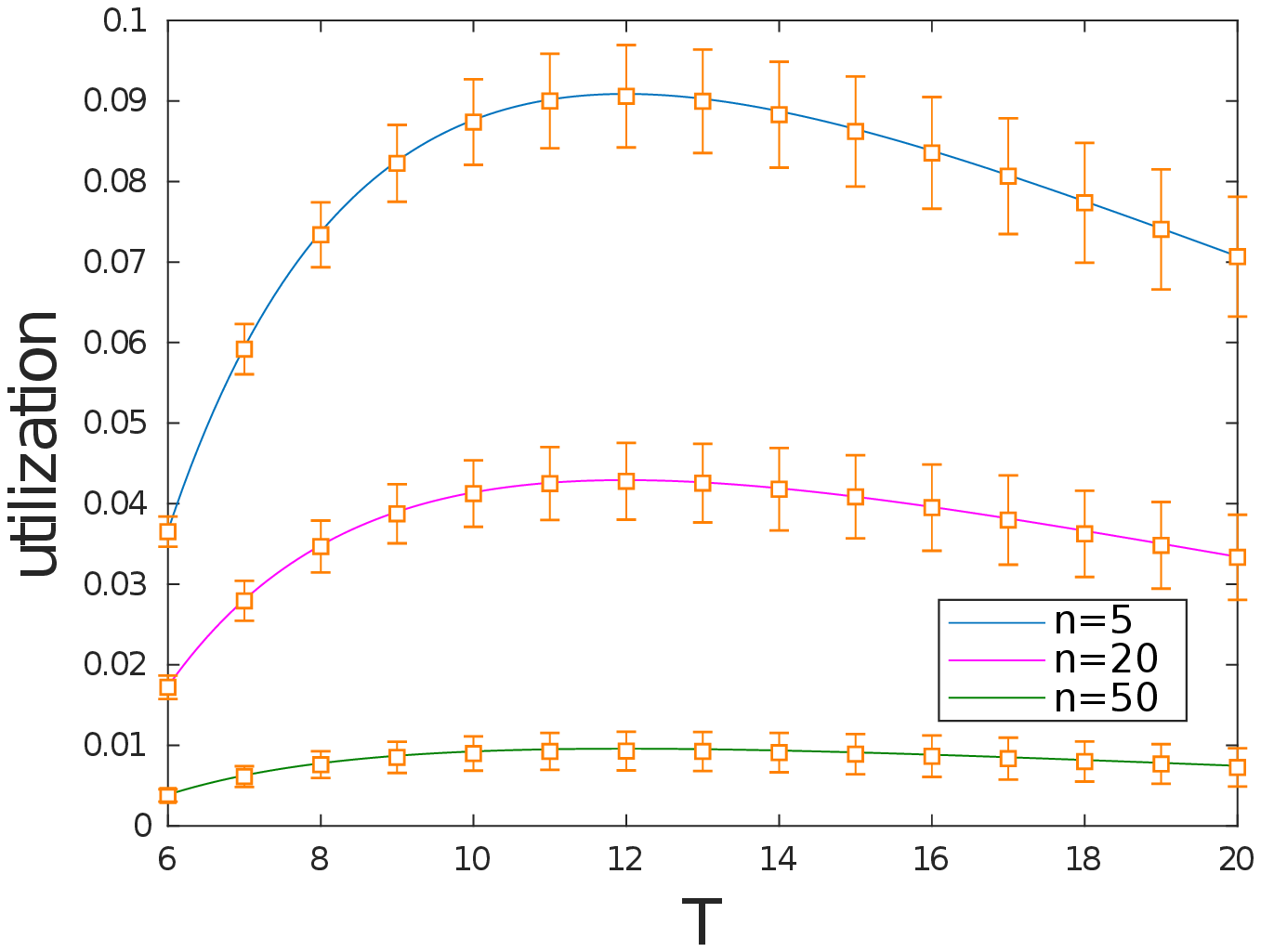}
        \caption{$\lambda$ = 0.1}    
        \label{fig:utiS5.1}
    \end{subfigure}
    \hfill
    \begin{subfigure}[b]{0.47\textwidth}  
        \centering 
        \includegraphics[width=2.5in, height=1.9in]{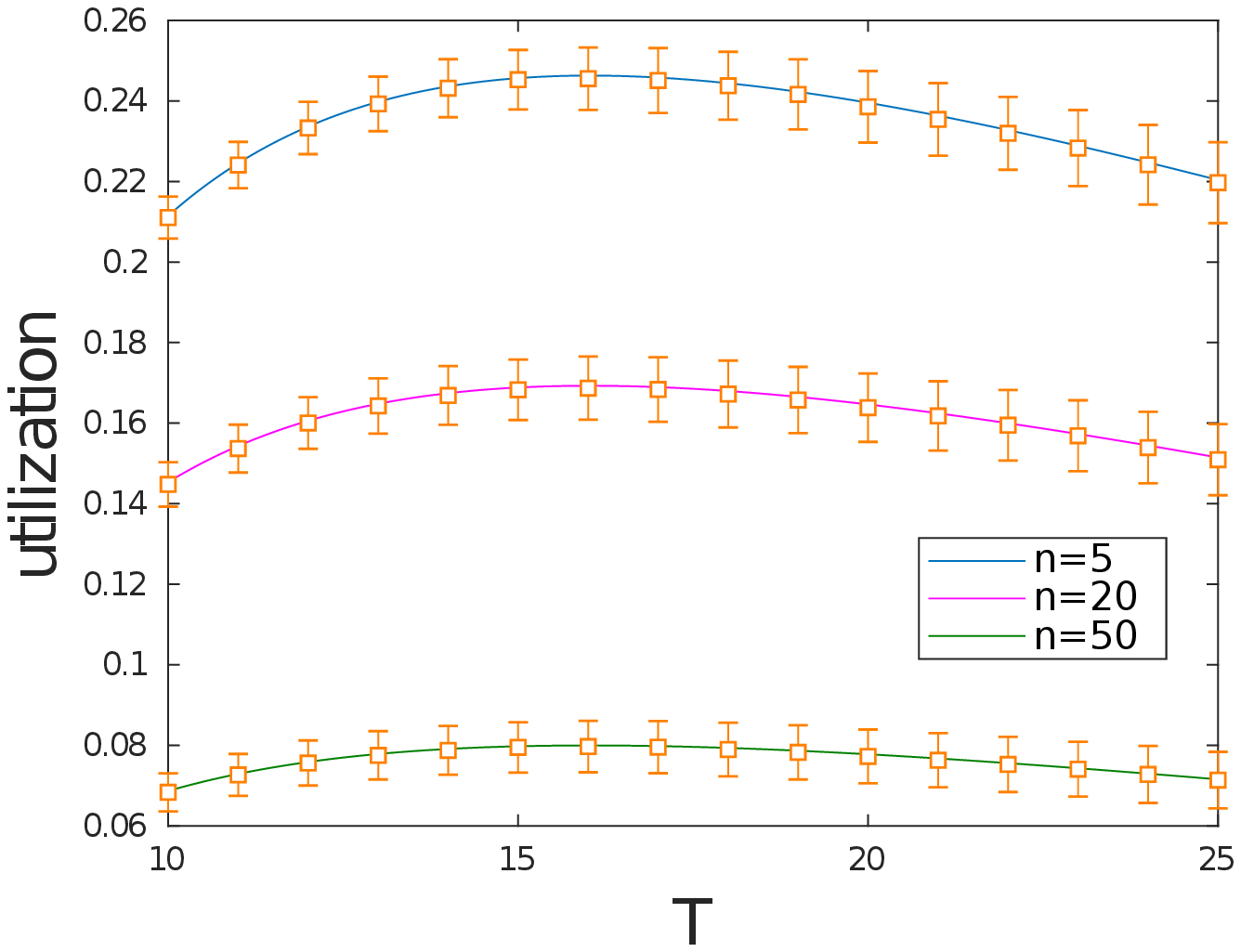}
        \caption{$\lambda$ = 0.05}     
        \label{fig:utiS5.05}
    \end{subfigure}
    \begin{subfigure}[b]{0.47\textwidth}   
        \centering 
        \includegraphics[width=2.5in, height=1.9in]{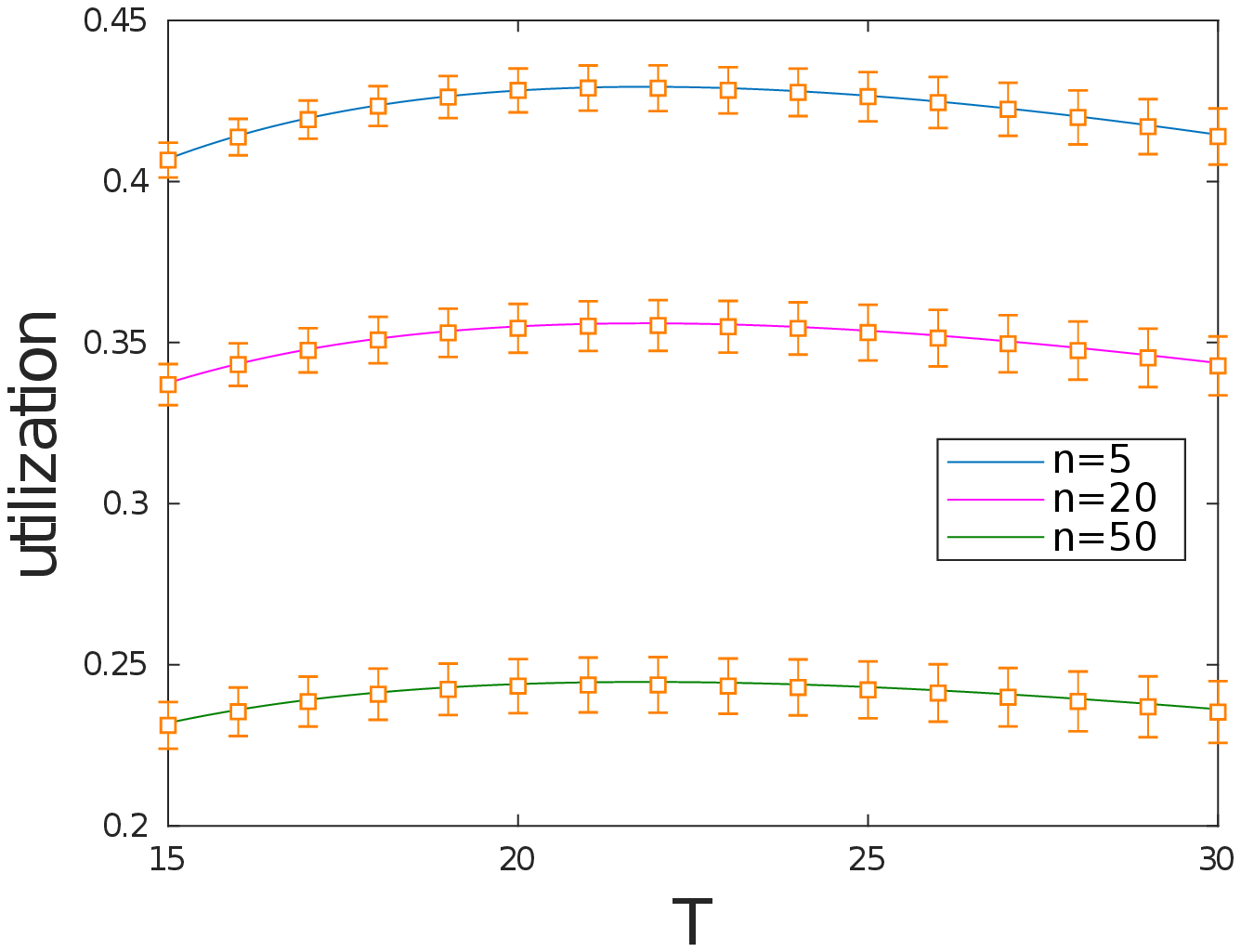}
        \caption{$\lambda$ = 0.025}     
        \label{fig:utiS5.025}
    \end{subfigure}
    \hfill
    \begin{subfigure}[b]{0.47\textwidth}
        \centering 
        \includegraphics[width=2.5in, height=1.9in]{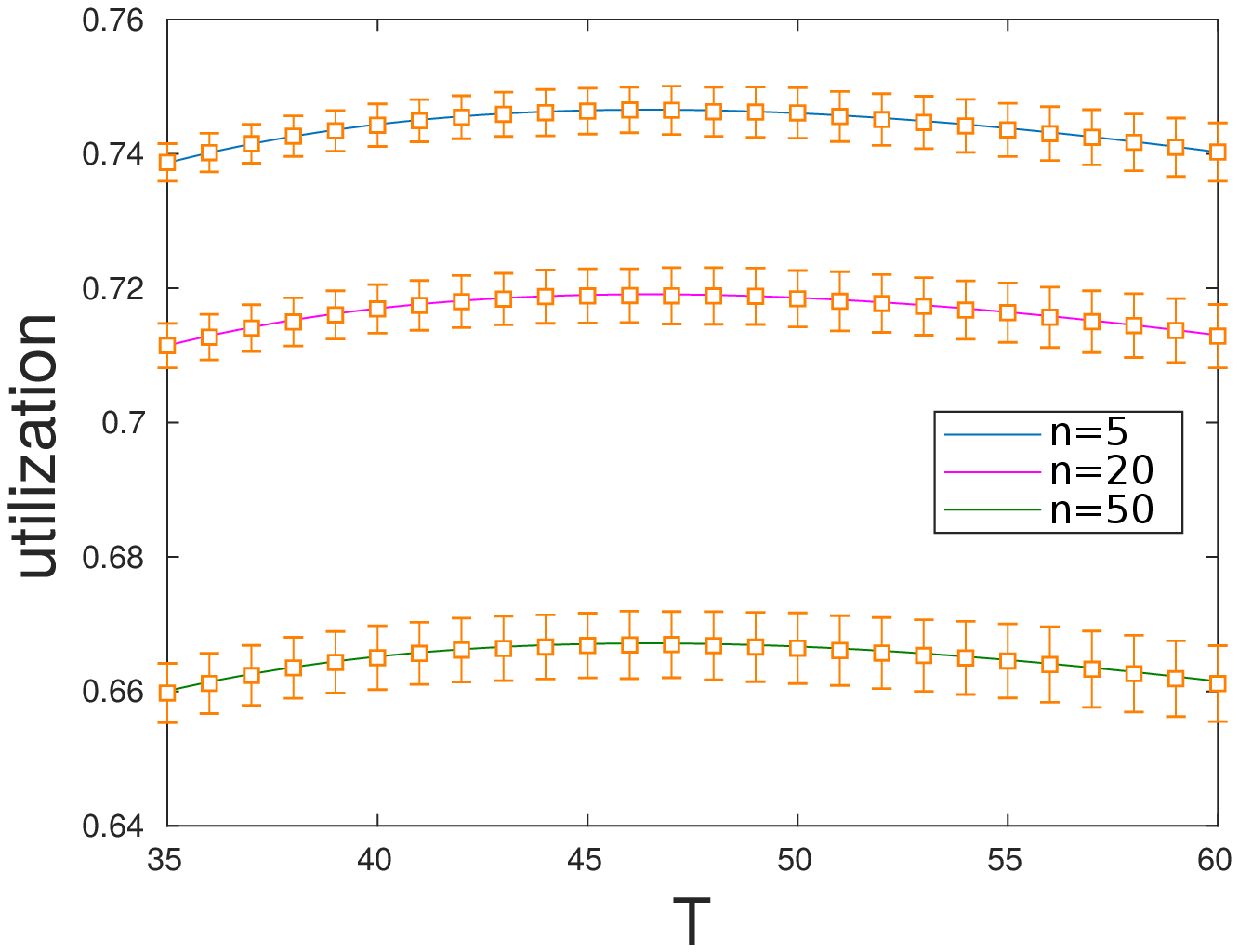}
        \caption{$\lambda$ = 0.005}
        \label{fig:utiS5.005}
    \end{subfigure}
    \caption{Utilization for $R=10$ minutes $c=5$ minutes and $\delta=0.5$ minutes. Solid lines represent theoretical utilization and data points with error bars represent the average utilization and the standard deviation.} 
    \label{fig:expUS50}
\end{figure}

Fig.~\ref{fig:expUS50} shows the utilization comparison between our model based on Eq.~\ref{eq:DAGU} and the simulation results for different failure rates using DAGs with different critical path lengths. The solid lines are theoretical utilization while the data points and error bars represent the average utilization and the standard deviation observed after 250 runs of the simulation, with each simulation running for $\frac{2000}{\lambda}$ minutes. For constant error rate, utilization decreases as $n$ increases and $T^*$ remains unchanged.

\section{Experimental Results with Apache Flink}
\label{sec:modelEvaluation}

We used Apache Flink, a state-of-the-art stream processing framework to evaluate the efficacy of our model. The purpose of this experiments is to show that our results accurately models existing implementations and it will improve over the default implementation parameters. While our experiments are not at large scale and involve some artificial failure injection, we believe that our experiments are still useful in this regard. We experimented with various Flink application instances that undertake word counting, with differing values of topology depth $n$, and we compared the observed utilization, $\tilde{U}$, obtained using the default Flink parameters for checkpoint interval, and other selected values of $T$, with our theoretical prediction of utilization, $U$, using measured parameters $\tilde{c}$, $\tilde{R}$ and $\tilde{\delta}$ from Flink's logs as inputs. We also set Flink's checkpoint interval to that given by the theoretical optimal, $T^*$, and observed the achieved utilization of the application. 

All experiments were conducted using \emph{m2.small} nodes running on OpenStack. Each virtual machine had one CPU core, 4 GB of RAM, 30 GB of disk space and ran Ubuntu 15.10, Java 1.7.0\_91, and Flink 1.3.3. We used a five node cluster with one node as master and four nodes as slaves. The word count application was loaded with a continuous stream of data and word counting was performed in a sliding window with a set of stateful operators to keep statistics of word counts in each window. We used the Hadoop Distributed File System as the state backend and used Apache Kafka consumer as the source to ensure that at a failure, all the records that were consumed between the last checkpoint and the failure were processed again. To simulate random failures, we killed one of the running Flink task managers based on an exponential distribution at precomputed failure event times. Using $\lambda$ values of 0.05, 0.01 and 0.005 per minute, we ran each experiment five times for 20, 30 and 40 hours respectively. These values of $\lambda$ are artificially large so as to indicate results that would be seen at a scale that we cannot experiment with due to lack of resources.

Table~\ref{tab:flinkeval} shows the settings of our experiments and observations: $\lambda$, $n$, observed $\tilde{c}$, $\tilde{R}$ and $\tilde{\delta}$, the observed utilization, $\tilde{U}$, and theoretical utilization, $U$, when $T=30$ minutes, the theoretical optimal, $T^*$, for the given settings and observed parameters, the observed and theoretical utilization when using the theoretical optimal, $T=T^*$, and the percentage increase in utilization $\%U$ over the default $T=30$ minutes. As shown in the table, we always observe utilization increase using the theoretical $T^*$. Moreover, as shown in the table theoretical predictions of utilization compare well to the observed utilization. In the table $T^*$ changes with a change in $n$ because $c$ changes with $n$, as a result of the windowing sizes changing with a deeper topology which increases the checkpoint cost.

\begin{table*}[t]
\centering
\renewcommand{\arraystretch}{1.2}
\setlength{\tabcolsep}{.1em}
\caption{Experimental results using Apache Flink}
\label{tab:flinkeval}
\begin{tabular}{||c|c|c|c|c|c|c||c|c|c|c||}
\hline
\multirow{2}{*}{$\lambda$} & \multirow{2}{*}{$n$} & \multirow{2}{*}{$\tilde{c}$ (s)} & \multirow{2}{*}{$\tilde{R}$ (s)}  & \multirow{2}{*}{$\tilde{\delta}$ (ms)}   & \multicolumn{2}{c||}{$T=30$ minutes} & \multirow{2}{*}{\shortstack{$T^*$ \\(min)}} & \multicolumn{2}{c|}{$T=T^*$} & \multirow{2}{*}{\shortstack{$\%U$ \\ increase}}\\
\cline{6-7}\cline{9-10}
 & & & & & $\tilde{U}$ & $U$ & & $\tilde{U}$& $U$ & \\
\hline
\multirow{2}{*}{.05} & 5 & 1.6$\pm$0.026 & 23.1$\pm$0.269  & 27.35$\pm$3  & .4228$\pm$.014 & .4222 & 1.0418 & .9416$\pm$.002 & .9442  & 123\% \\
& 7 & 3.09$\pm$0.079 & 23.81$\pm$0.22  & 32.65$\pm$2  & .4069$\pm$.011 & .4216 & 1.4526 & .9223$\pm$.002 & .9258 & 126\% \\
\hline
\multirow{2}{*}{.01}& 5 & 1.07$\pm$0.022 & 23.70$\pm$1.095  & 13.86$\pm$1  & .8449$\pm$.001 & .8536 & 1.8945 & .9806$\pm$.001 & .98 & 16\% \\
& 7 & 1.59$\pm$1.441 & 24.12$\pm$0.438  & 15.07$\pm$1 & .8447$\pm$.001 & .8533 & 2.3110 & .9686$\pm$.001 & .9692 & 15\% \\
\hline
\multirow{2}{*}{.005}& 5 & 1.15$\pm$0.427 & 25.37$\pm$1.064  & 12.6$\pm$1  & .9109$\pm$.005 & .9243 & 2.7753 & .9815$\pm$.005 & .9866 & 8\% \\
& 7 & 2.57$\pm$0.117 & 24.07$\pm$0.411  & 12.85$\pm$1  & .9118$\pm$.001 & .9237 &  4.1536 & .9755$\pm$.001 & .9781 & 7\% \\
\hline
\end{tabular}
\end{table*}

We also calculated the percentage utilization increase achieved using the theoretical optimal compared to the default 30 minute checkpoint interval, when $R=30s, c=5s, \delta=50ms$ and $n=5$, for the five real-world distributed systems in~\cite{8549548} that have failure rates 0.8475, 0.1701, 0.135,  0.1161 and 0.0606 per hour. In this case we can achieve utilization increase of 18.91\%, 2.4\%, 1.73\%, 1.4\% and 0.5\% respectively. The default 30 minute checkpoint interval gives the optimal utilization for failure rate, $\lambda=   0.0022$ per hour assuming $c=1s$. Most distributed systems have higher failure rates than 0.0022 per hour indicative of roughly 1 failure every 19 days. Therefore, using 30 minute checkpoint interval in real-world systems can lead to inefficient system operation.

\subsection{Comparison and scaling up}

\begin{figure}[h]  
   \centering
		\includegraphics[width=2.9in]{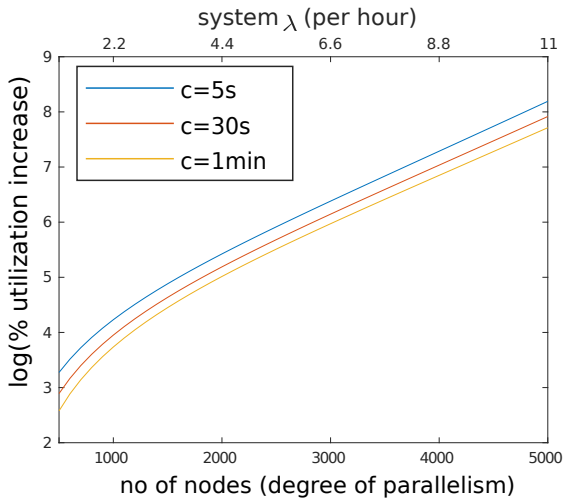}
   \caption{System failure rate and the utilization increase gained using $T^*$ instead of 30 minutes for increasing number of nodes.} 
   \label{fig:clusterSize}
   \end{figure}

Considering the scaling up of distributed stream processing systems, the failure rate of the system increases with the number of nodes in the system. For example, in Exascale systems multiple failures are expected everyday~\cite{5871590,Bianchini2009SystemRA} and MTTF is anticipated to be in minutes~\cite{phdthesis,doi:10.1177/1094342009347767}. For systems such as Flink where a failure of a single node results in restarting the whole application from the previous checkpoint, the failure rate of the system is $\sum_{i=1}^{n} \lambda_i$, where $\lambda_i$ is the failure rate of node $i$~\cite{1303239}. 
Fig.~\ref{fig:clusterSize} shows how the failure rate changes with the number of nodes in the system considering the failure rate of all nodes is 0.0022 per hour. The figure also shows the percentage utilization increase we obtain using $T^*$ instead of using 30 minute checkpoint interval for $R=30s, c=5s, \delta=50ms$ and $n=5$. As the number of nodes increases, failure rate increases and the utilization increase we obtain also increases significantly. For example, as indicated in the figure, we can achieve 68.8\% utilization increase for 1000 nodes and 226.83\% for 2000 nodes. Furthermore, as the checkpointing cost increases, percentage utilization increase reduces as well.

\begin{figure}[h]  
   \centering
	\includegraphics[width=3in]{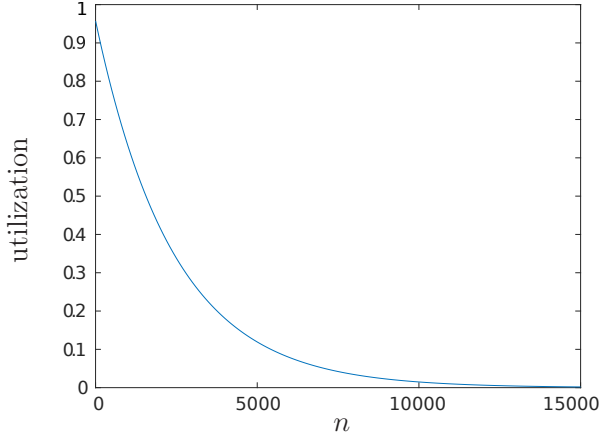}
   \caption{Decrease in utilization with the increase of $n$ using $T^*$.} 
   \label{fig:nVsU}
   \end{figure}

Utilization of a streaming application decreases with the increase of $n$. Although real-world streaming applications can be represented using a considerable low $n$ value, for applications with a large $n$ value, utilization will have a significant impact due to $\delta$ and $n$. Fig.~\ref{fig:nVsU} indicates how the utilization decreases for large $n$ values. The figure shows the utilization using the model expressed in Eq.~\ref{eq:DAGU} for $T^*,$ $R=30s, c=10s, \delta=5s$ and $\lambda=0.005$ per minute. As indicated in the figure for large $n$, the utilization comes close to 0. For example, in the figure utilization is 0.0018 for $n=15000$.
Therefore, large-scale applications with large $n$ and $\delta$ values can incur significant overheads due to the checkpointing process. 

\begin{figure}[h]   
\begin{subfigure}[b]{0.49\textwidth}
        \centering
        \includegraphics[width=2.32in]{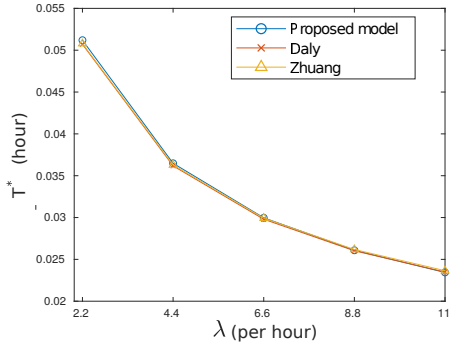}
    \caption{$c=10$ seconds and $R=30$ seconds.} 
    \label{fig:compare1}
    \end{subfigure}
    \hfill    
    \begin{subfigure}[b]{0.49\textwidth}  
        \centering 
        \includegraphics[width=2.3in]{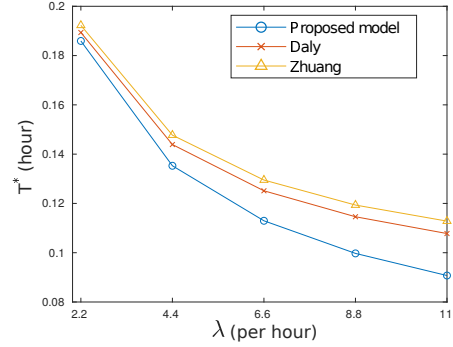}
   \caption{$c=2$ minutes and $R=5$ minutes.} 
   \label{fig:compare2}
    \end{subfigure}
     \caption{Optimal checkpoint interval comparison of the proposed $T^*$, Daly's model~\cite{Daly:2003:MPO:1757599.1757601} and model of Zhuang et al.~\cite{8487327}.} 
     \label{fig:compare}
\end{figure}

\begin{figure}[h]  
   \centering
   \includegraphics[width=2.3in]{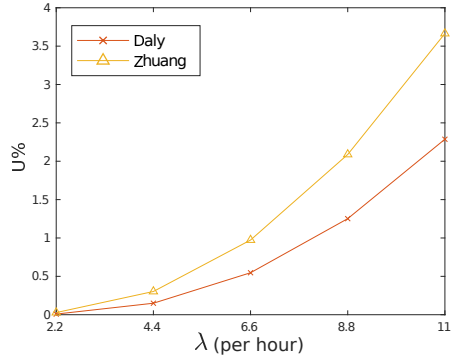}
   \caption{Percentage increase in utilization using the proposed $T^*$ instead of Daly's model~\cite{Daly:2003:MPO:1757599.1757601} and model of Zhuang et al.~\cite{8487327} for $c=2$ minutes, $R=5$ minutes, $\delta=30$ seconds and $n=25.$} 
   \label{fig:Ucompare}
   \end{figure}

Fig.~\ref{fig:compare} shows the values of $T^*$ given by our model, Daly's first order model~\cite{Daly:2003:MPO:1757599.1757601}, $\sqrt{2c(\frac{1}{\lambda}+R)}$ and Zhuang et al.'s model~\cite{8487327}, $\sqrt{2c(\frac{1}{\lambda}+R)+c^2}$. For the model of Zhuang et al., we assumed maximum processing rate and the average input rate of the system is same. 
All the models give near similar results for smaller $c$ and $R$ values as shown in Fig.~\ref{fig:compare1}. However, as the values of $c$ and $R$ get bigger, our model deviates from the other models for larger $\lambda$ values. This is because the assumptions made in Daly's first order model are not accurate for large $\lambda$  values as stated in Daly's paper~\cite{Daly:2003:MPO:1757599.1757601}. Moreover, we calculated the utilization based on Eq.~\ref{eq:DAGU} for the proposed $T^*$ and the optimal values given by Daly and Zhuang et al. presented in Fig.~\ref{fig:compare2} for $\delta=30$ seconds and $n=25.$  Fig.~\ref{fig:Ucompare} shows the percentage utilization increase gained using the proposed $T^*$ compared to other models. As indicated in the figure, the utilization increase we can gain using the proposed model becomes more significant as $\lambda$ increases. For example, when $\lambda=11$, the utilization increase gained using the proposed $T^*$ compared to the values given by Daly's model and Zhuang~et al.'s model is $2.3\%$ and $3.7\%$ respectively.

% \subsection{Other considerations}
% 
% Real-world systems use heartbeat to detect failures and therefore, the heartbeat interval of a system affects the value of $R$, the time to detect and recover from a failure. If the heartbeat interval is $H$ and given that a failure happens during time $H$, then we know from Eq.~\ref{eq:meantime} that the mean time to failure is $\mathrm{F}(H)$. Therefore, the average time to detect failures is $H-\mathrm{F}(H)$. Taking all of this into account, we can write $R = (H-\mathrm{F}(H)) + recovery \ time,$ where recovery time is the time taken to load the last persisted checkpoint.
% 
% In stream processing systems, the input rate can vary from time to time which affects the time required to redo the work lost due to failures. We can further improve the model by taking this factor into account. $\mathrm{F}(T)$ indicates the time lost if a failure happens and the amount of rework to be done which was lost during $\mathrm{F}(T)$ depends on the input rate, $i$ and processing rate, $p$ of the system. For example, if the input rate is very low between the last checkpoint and the failure, then the time taken to redo the lost computations could be much less than, $\mathrm{F}(T)$.  Therefore, we can improve the rework time as $\mathrm{F}(T)i/p$.

\section{Conclusion}\label{sec:conclusion}
We provided a rigorous analytical expression for the utilization of a distributed stream processing system that allows optimization of the checkpoint interval through maximizing utilization. The optimal checkpoint interval is seen to be dependent only on checkpoint cost and failure rate. Our analytical formulation provides a solid theoretical basis for the analysis and optimization of more elaborate checkpointing approaches such as \emph{multi-level checkpointing}~\cite{5645453}; where Moody et al. show that considering a hierarchy of faults, or multi-level failure model, provides improvement. 

In stream processing systems, the input rate can vary widely which can affect the checkpoint cost, e.g. if no data is observed for some time the checkpoint cost may decrease significantly. As $T^*$ depends only on $c$ and $\lambda$, in each checkpoint interval we may measure the $c$ of completed checkpoint and $\lambda$, and update $T^*$ dynamically for the next checkpoint interval, to adapt to changing workloads; which we leave to future work.

\section*{Acknowledgment}
This research is funded in part by the Defence Science and Technology Group, Edinburgh, South Australia, under contract MyIP:6104, and was supported by use of the Nectar Research Cloud, a collaborative Australian research platform supported by the National Collaborative Research Infrastructure Strategy (NCRIS).

\bibliographystyle{ACM-Reference-Format}
\bibliography{preprint}
\end{document}